\documentclass[11pt,a4paper]{article}
\usepackage[utf8]{inputenc}
\usepackage[T1]{fontenc}
\usepackage{lmodern}
\usepackage{microtype}
\usepackage[margin=2.5cm]{geometry}
\usepackage{setspace}
\usepackage{parskip}
\usepackage{wrapfig}
\usepackage{graphicx}
\usepackage{booktabs}
\usepackage{enumitem}
\usepackage{hyperref}
\hypersetup{colorlinks=true,linkcolor=blue,citecolor=blue,urlcolor=blue}
\usepackage{titling}
\usepackage{fancyhdr}
\usepackage{caption}
\usepackage{amsmath,amssymb}

\pretitle{\vspace*{-1cm}\begin{center}\LARGE\bfseries}
\posttitle{\par\end{center}\vskip 0.5em}
\preauthor{\begin{center}\large}
\postauthor{\end{center}}
\predate{\begin{center}\large}
\postdate{\end{center}}

\begin{document}

\begin{titlepage}
  \centering
  \vspace*{1cm}
  {\Huge\bfseries Expanding Horizons \\[6pt] \Large Transforming Astronomy in the 2040s \par}
  \vspace{1.5cm}

  {\LARGE \textbf{The supernova Ia progenitor problem in the 2040s}\par}
  \vspace{1cm}

  \begin{tabular}{p{4.5cm}p{10cm}}
    \textbf{Scientific Categories:} & compact binaries, compact objects, type Ia supernovae, binary evolution \\
    \\
    \textbf{Submitting Author:} & Name: Thomas Kupfer \\
    & Affiliation: University of Hamburg \\
    & Email: thomas.kupfer@uni-hamburg.de \\
    \\
    \textbf{Contributing authors:} & Simone Scaringi (Durham University, UK)
Ingrid Pelisoli (University of Warwick, UK)
Anna F. Pala (ESO, DE)
Silvia Toonen (University of Amsterdam, the Netherlands)
Domitilla de Martino (INAF-OACNa, IT)
Christa Gall (University of Copenhagen, DK)
Kunal Deshmukh (KU Leuven, BE)
Valerie Van Grootel (University of Liège, BE)
Stéphane Blondin (ESO, DE and LAM, FR)
Samaya Nissanke (DESY, DZA, University of Potsdam, DE; Amsterdam, NL)
 \\
  \end{tabular}

  \vspace{1cm}

  \textbf{Abstract:}

  \vspace{0.5em}
  \begin{minipage}{0.9\textwidth}
    \small 
    Type Ia supernovae (SNe Ia) are fundamental to cosmology and galactic chemical evolution, yet the nature of their progenitor systems remains unresolved. Multiple evolutionary pathways—including single-degenerate, double-degenerate, and helium-donor systems—are thought to contribute to the SN Ia population, but direct observational constraints are limited. This uncertainty hampers our understanding of SN Ia diversity and introduces systematic uncertainties in their use as precision cosmological probes. By the 2040s, surveys such as Gaia, LSST, SDSS-V, 4MOST, and the gravitational-wave mission LISA will identify thousands of compact binaries in the Milky Way that are potential SN Ia progenitors. However, survey discoveries alone are insufficient. Robust identification and characterization require high–time-resolution, phase-resolved spectroscopy to determine fundamental parameters such as component masses, orbital inclinations, chemical compositions, and accretion states. Addressing these challenges demands new observational capabilities. The most compact binaries require continuous, dead-time–free spectroscopy with negligible readout noise, while the progenitor population spans a wide range of brightness and orbital periods. A modular, multi-aperture telescope array equipped with fast, low-noise spectrographs can flexibly combine collecting area for faint targets, observe bright systems efficiently, and deliver uninterrupted time series through staggered exposures. Such observations are difficult for single-aperture facilities.

  \end{minipage}

\end{titlepage}


\section*{Introduction}
Type Ia supernovae (SNe Ia) play a central role in modern astrophysics. Their remarkable uniformity has made them the cornerstone of observational cosmology, enabling the discovery of the accelerated expansion of the Universe (Riess et al. 1998; Perlmutter et al. 1999). SNe Ia are also major contributors to the enrichment of iron-peak elements in galaxies and are therefore crucial for chemical evolution of the universe (Matteucci \& Greggio 1986, Matteucci 2012, 2021). Despite their importance, the nature of the stellar systems that give rise to SNe Ia remains one of the most significant unresolved questions in astrophysics. Although only the thermonuclear explosion of a carbon-oxygen white dwarf (WD) following interaction with a binary companion can explain the observed features, much less is known about their progenitors and relation to the cause of the explosion. The lack of a definitive progenitor model limits our understanding of the diversity of SNe Ia, the precision of cosmological measurements, and the interpretation of chemical evolution signatures across cosmic time.

Two broad families of progenitor channels dominate current discussions: the single-degenerate scenario, where a white dwarf accretes from a non-degenerate companion (Whelan \& Iben 1973), and the double-degenerate scenario, involving the merger of two white dwarfs (Webbink 1984; Iben \& Tutukov 1984). Additional channels such as helium-donor systems leading to double-detonation explosions have gained support from both theory and observation as well (e.g. Fink et al. 2010; Woosley \& Kasen 2011). Observations have shown that SNe Ia show a large range of explosion energies and decay times, photometric and spectroscopic signatures indicating different progenitor systems (e.g. Röpke et al. 2012, Jha et al. 2019 and references therein). Recently, Das et al. (2025) found evidence for the double detonation scenario in a supernova remnant. However, the number of systems with WDs massive enough to be SN Ia progenitors is tiny: only three confirmed systems for the double degenerate channel (Geier et al. 2007, Pelisoli et al. 2021, Munday et al. 2025) and two candidates for the double detonation channel (Geier et la. 2013, Kupfer et al. 2022). 

The difficulty in identifying progenitor systems stems from several challenges: their intrinsic faintness, the diversity of evolutionary histories, and the short-lived nature of the most relevant evolutionary phases. Additionally, much of the structure and dynamics of these binaries can only be constrained through high-precision, phase-resolved spectroscopy. Current facilities allow this only for the brightest nearby systems, leaving the vast majority of candidates beyond our observational reach. These challenges are especially acute for the single-degenerate channel, where a faint white dwarf must be detected next to a far more luminous non-degenerate companion. Detecting the WD contribution requires high sensitivity at short wavelengths—where the white dwarf produces a detectable blue or UV excess—and careful decomposition of the spectral energy distribution. For donors in wider orbits, such as subgiant or red-giant companions, the long periods demand extended radial-velocity monitoring and sometimes observations at specific orbital phases to detect subtle accretion signatures or reflex motion. As we approach the 2040s, a new generation of facilities will be required to overcome these limitations and open previously inaccessible regions of parameter space. This white paper outlines the key scientific questions that will shape SN Ia research in the coming decades and describes the capabilities needed from a next-generation ESO instrument designed to revolutionize our understanding of SN Ia progenitors.

\section*{Open Science Questions for the 2040s: The Progenitor Problem in a New Era}
By the 2040s, ongoing and future surveys such as Gaia, VRO (LSST), 4MOST, BlackGEM, SDSS-V, and space-based missions like LISA will have dramatically expanded the census of compact binaries in the Milky Way. Gaia has already revealed thousands of white dwarfs in binary systems; LSST will provide variability and eclipsing information for millions of faint sources; LISA will detect tens of thousands of compact binaries through gravitational waves (Nelemans et al. 2001; Amaro-Seoane et al. 2023). This revolution in data volume will not by itself solve the progenitor puzzle. Instead, it will highlight the gaps that only time-resolved spectroscopy at medium resolution can fill. As we move into the 2040s, several key scientific questions will remain open.

A central issue is the relative contribution of different progenitor channels to the overall SN Ia population. Cosmological analyses require precise knowledge of the intrinsic diversity of SNe Ia and of how explosion properties depend on progenitor type. Yet binary population synthesis models predicting the frequency of single-degenerate, double-degenerate, and helium-donor scenarios remain below observed populations (e.g. Ruiter et al. 2009), primarily because their input physics—mass transfer efficiency, common-envelope outcomes, and binary angular momentum loss—are poorly constrained observationally. A large, systematically observed sample of progenitor candidates spanning a wide range of luminosities, metallicities, and evolutionary stages is essential for resolving these uncertainties.

Another open question concerns the existence and characteristics of sub-Chandrasekhar mass explosions. Theoretical work has shown that detonations in white dwarfs below the Chandrasekhar mass may reproduce many observed SN Ia light curves and spectra (Sim et al. 2010; Shen et al. 2018), but direct progenitor analogues remain sparse. Identifying systems in which a white dwarf accretes helium from a hot subdwarf or low-mass helium star is key to testing these models. Importantly, such systems do exist: several binaries with hot subdwarf (sdB) donors transferring matter to a white dwarf have already been identified (e.g. Kupfer et al. 2020), and additional candidate progenitors—representing different stages of the helium-donor pathway—are also known.

A third frontier emerging in the 2040s is the integration of gravitational-wave observations with multiwavelength electromagnetic data. LISA will detect compact white-dwarf binaries that are otherwise invisible, enabling a new pathway to identify potential SN Ia progenitors through their orbital decay signatures. However, gravitational-wave data alone cannot determine component masses, chemical compositions, or evolutionary states. Only coordinated electromagnetic observations—especially phase-resolved spectroscopy—can provide the necessary system parameters. This synergistic era will significantly expand the dynamic range of accessible progenitor systems but will require instrumentation capable of studying sources that are an order of magnitude fainter than what today’s large telescopes can handle.

Finally, there remains the outstanding question of environmental dependence. Metallicity is known to affect stellar winds, mass-loss rates, and therefore binary evolution. Determining how progenitor channels vary across stellar populations in the thin disk, thick disk, bulge, and halo is essential for interpreting extragalactic SN Ia statistics and for accurately reconstructing chemical enrichment histories. Addressing this requires the ability to characterize progenitor systems across a wide range of stellar environments, including regions where foreground reddening and low brightness currently impose severe limitations.

\begin{figure*}
   \centering
   \includegraphics[width=0.82\hsize]{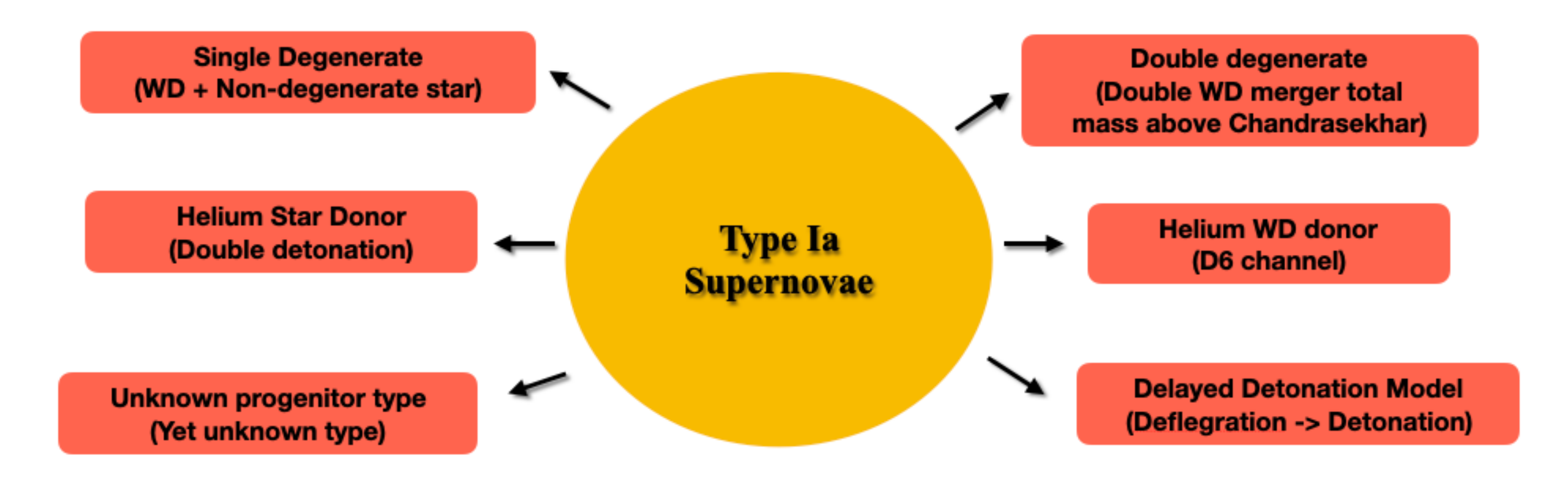}
   \caption{Overview of the likely dominant channels discussed in the literature}
   \label{Corner}
\end{figure*}

\section*{Technology and data handling requirements}
Resolving the nature of SN Ia progenitors in the 2040s requires a new class of instrumentation capable of obtaining high–time-resolution, phase-resolved spectra at medium resolution covering the full spectral range from UV to NIR across the full diversity of compact binaries. Existing large telescopes are limited by detector readout time, readout noise, and inflexible observing modes, making them unsuited for characterizing the shortest-period systems—particularly the ultracompact binaries that will be discovered by LISA.
A next-generation facility must therefore provide continuous, dead-time–free spectroscopy with zero readout noise to capture rapid radial-velocity and line-profile variations. This capability is essential for systems with orbital periods of only a few minutes, where even short interruptions degrade phase coverage and prevent accurate mass and radius determinations.
To achieve this, we envision an array of telescopes, each equipped with fast, low-noise, medium to high resolution spectrographs. Operating as a modular system, such an array can dynamically match observational demands:

\begin{itemize}
    \item combine multiple apertures for faint, ultracompact binaries
\item operate individual telescopes at high spectral resolution for bright nearby systems
\item provide long-cadence monitoring for symbiotic binaries with giant donors, and 
\item use medium to high resolution modes to measure precise system parameter such as radial velocities, effective temperatures, surface gravities, and abundances
\end{itemize}

By staggering exposure start times across the array, the facility can deliver a truly continuous spectroscopic time series—something fundamentally impossible with a single-aperture telescope, including the ELT. Flexible scheduling and adjustable cadence are critical to follow both the fastest LISA sources and slower, mass-transferring supernova Ia progenitors.

\section*{References}
\small{Amaro-Seoane, P., et al. 2023, Living Reviews in Relativity, 26, 1 - Das, P., et al. 2025, ApJ, submitted - Fink, M., et al. 2010, A\&A, 514, A53 - Geier, S., et al. 2007, A\&A, 464, 299 - Geier, S., et al. 2013, A\&A, 554, A54 - Iben, I., \& Tutukov, A. V. 1984, ApJS, 54, 335 - Jha, S. W., Maguire, K., \& Sullivan, M. 2019, Nature Astronomy, 3, 706 - Kupfer, T., et al. 2020, ApJ, 891, 45 - Kupfer, T., et al. 2022, ApJ, 926, L 8- Matteucci, F. 2012, Chemical Evolution of Galaxies (Springer) - Matteucci, F. 2021, A\&A Rev., 29, 5 - Matteucci, F., \& Greggio, L. 1986, A\&A, 154, 279 - Munday, J., et al. 2025, NatAst, 9, 872 - Nelemans, G., et al. 2001, A\&A, 365, 491 - Perlmutter, S., et al. 1999, ApJ, 517, 565 - Pelisoli, I., et al. 2021, Nature Astronomy, 5, 1052 - Riess, A. G., et al. 1998, AJ, 116, 1009 - Röpke, F. K., et al. 2012, ApJ, 750, L19 - Ruiter, A. J., et al. 2009, ApJ, 699, 2026 - Shen, K. J., Bildsten, L., Kasen, D., \& Quataert, E. 2018, ApJ, 865, 15 - Sim, S. A., Röpke, F. K., et al. 2010, ApJL, 714, L52 - Webbink, R. F. 1984, ApJ, 277, 355 - Whelan, J., \& Iben, I. 1973, ApJ, 186, 1007 - Woosley, S. E., \& Kasen, D. 2011, ApJ, 734, 38}

\end{document}